\documentclass{article}

\usepackage{microtype}
\usepackage{booktabs}

\usepackage{hyperref}
\usepackage[accepted]{icml2024}  

\makeatletter
\renewcommand{\ICML@appearing}{\textit{Accepted at the ICML 2026 Workshop on
AI Forecasting (non-archival).}}
\makeatother

\hypersetup{
  pdftitle={Leakage-Aware Benchmarking of LLM Forecasting: Real-Time Nowcasts as the Decision-Time Input for Macro Factor Ranking},
  pdfauthor={Mao Guan, Qian Chen},
}

\usepackage{url}
\usepackage{amsmath}
\usepackage{amssymb}
\usepackage{enumitem}
\usepackage{xcolor}
\usepackage{tikz}
\usepackage{graphicx}
\usepackage{float}
\usetikzlibrary{positioning,arrows.meta,shapes.geometric,fit,backgrounds}

\icmltitlerunning{Leakage-Aware Benchmarking of LLM Forecasting}

\begin{document}

\twocolumn[
\icmltitle{Leakage-Aware Benchmarking of LLM Forecasting:\\
Real-Time Nowcasts as the Decision-Time Input for Macro Factor Ranking}

\begin{icmlauthorlist}
\icmlauthor{Mao Guan}{ind}
\icmlauthor{Qian Chen}{ind}
\end{icmlauthorlist}

\icmlaffiliation{ind}{Independent Researcher}

\icmlcorrespondingauthor{Mao Guan}{guanmao771@gmail.com}

\icmlkeywords{forecasting, real-time data, publication lag, large language
models, retrieval-augmented generation, financial forecasting, factor ranking,
benchmark}

\vskip 0.3in
]

\printAffiliationsAndNotice{}

\begin{abstract}
Forecasting benchmarks for retrieval-augmented LLMs routinely confound
model capability with information leakage: features labeled with a
target's timestamp are often not observable at the system's decision
time. We study leakage-controlled equity factor ranking with a
retrieval-augmented 7B open-source LLM forecaster. At each month-end from
2023-04 to 2026-03, the forecaster observes only decision-time information:
lag-shifted FRED macro variables, recent macro-event summaries, and the
Cleveland Fed's archived daily CPI nowcast for unreleased current-month
inflation. A macro-analog retrieval module selects historical states, a critic
LLM compresses them into one tactical rule, and an actor LLM maps the current
state and recent rules into scores for seven U.S. equity style factors. The
full pipeline obtains a median monthly Spearman rank IC of $+0.154$, with
positive means across three non-overlapping contiguous 12-month subwindows;
the mean IC remains statistically underpowered, with a bootstrap 95\%
confidence interval that includes zero. Non-LLM baselines under the same
decision-time constraint demonstrate that a kNN macro-analog model recovers a
comparable median IC, indicating that real-time inflation information and
macro-similar retrieval explain much of the median signal. The LLM pipeline
retains higher mean IC and a stronger long-short allocation sanity check,
suggesting that any marginal benefit is concentrated in the extreme
rankings that drive long-short portfolio formation. A descriptive audit of
the 36 critic rules and per-month case studies appears in the appendix.
\end{abstract}

\section{Introduction}\label{sec:intro}

Forecasting systems are often evaluated under a fiction of real-time
observability. In macro-driven financial forecasting, this fiction is
consequential: the CPI value labeled as month $t$ is typically not available
at the month-end decision date for month $t$: the U.S. Bureau of Labor Statistics (BLS) publishes it about ten
days later. A walk-forward backtest that reads this value directly leaks
future information~\citep{bailey2014deflated}.

This issue is amplified in LLM and retrieval-augmented forecasting
pipelines~\citep{lopezlira2023chatgpt,xiao2024tradingagents}, and recent
benchmarks have begun to evaluate LLM-based agents on financial
decision-making tasks~\citep{li2025investorbench}.
Such systems ingest macro narratives, event summaries, and historical analogs,
but these inputs are only meaningful if they respect what a decision-maker
could have observed at the time. We therefore ask whether a 7B
retrieval-augmented LLM forecaster can rank equity style factors under a
strict decision-time information constraint, and how much of any observed
signal is attributable to the LLM architecture versus to the underlying
decision-time information set~\citep{chen2025standardbenchmarks}.

We build a leakage-controlled monthly benchmark for ranking seven U.S. equity
style factors. The key design choice is to replace unreleased current-month
CPI with the Cleveland Fed's archived daily inflation nowcast~\citep{knotek2017nowcasting, clevelandfednowcast}, while shifting FRED CPI and unemployment by one
month. A macro-analog retrieval module selects historical states with a strict
$\geq 12$-month embargo, a critic LLM writes one tactical rule from the
analogs~\citep{shinn2023reflexion}, and an actor LLM maps the current real-time state into factor scores.
The 2023-04 to 2026-03 evaluation window is set by the joint availability of
factor returns, lag-corrected FRED macro inputs, the Cleveland Fed nowcast
archive, and the event corpus.

Our contributions are fourfold. \textit{(i)} We construct a
leakage-controlled walk-forward benchmark for macro-conditioned equity factor
ranking that respects publication lags throughout~\citep{harvey2016cross}. \textit{(ii)} We introduce
a macro-RAG actor-critic forecaster that compresses retrieved historical
analogs into tactical rules before scoring factors, runnable with 4-bit
quantization on commodity GPU hardware. \textit{(iii)} Under the same
decision-time information set, we disentangle the contributions of model
class and decision-time information: a kNN macro-analog baseline recovers a
comparable median rank IC to the full LLM pipeline, while the LLM's
residual edge concentrates in mean IC and a long-short allocation sanity
check that are driven by extreme-ranking months and do not reach
conventional significance at $n=36$. \textit{(iv)} We release the 36 reproducible critic rules,
the parsed Cleveland Fed nowcast archive frozen on 2026-05-13, and
representative case studies, alongside the walk-forward harness.

\section{Setup}\label{sec:setup}

\textbf{Factors.} Seven cross-sectional U.S. equity style factors: SMB, HML,
RMW, CMA~\citep{fama2015five}, UMD~\citep{jegadeesh1993returns}, BAB~\citep{frazzini2014betting}, and QMJ~\citep{asness2019quality}. Sources: the
Kenneth French Data Library~\citep{frenchdatalibrary} and the AQR Capital
Management public factor library~\citep{aqrdatasets}. The market factor
(MKT\_RF) is excluded from prediction.

\textbf{Label.} The label at month $t$ is the cross-sectional rank of the
market-adjusted residual return $r^{\text{resid}}_{f,t} := r_{f,t} -
\beta_{f,t}\,r_{\text{MKT},t}$, where $\beta_{f,t}$ is the rolling 60-month
OLS beta estimated strictly with data $s < t$.

\textbf{Decision-time inputs.} All methods use decision-time-clean inputs:
FRED CPI YoY, CPI level, and Unemployment shifted by one calendar month, plus
the daily Cleveland Fed CPI YoY nowcast queried at date $t$. All remaining
market variables from FRED, such as term spread and VIX, are treated as
observable at $t$ without publication lag. The non-LLM baselines use only
this structured macro / nowcast subset; the LLM actor additionally receives
decision-time event summaries from recent FOMC and CPI releases.

\textbf{Walk-forward.} 36 monthly decisions, 2023-04-30 through 2026-03-31.

\textbf{Metric.} Per-month Spearman rank IC between the model's seven scores
and the realised seven ranks of $r^{\text{resid}}_{f,t}$. We report mean,
median (outlier-robust), three contiguous 12-month sub-window means, bootstrap
95\% CI on the mean (10{,}000 resamples, seed $42$), and a sign-symmetric
permutation $p$-value (10{,}000 permutations). With $n=36$ and per-month IC
standard deviation $\approx 0.45$ (close to the Spearman null when ranking
seven items), single-window tests have low statistical power at the 5\%
level for any modest signal.

\section{Method}\label{sec:method}

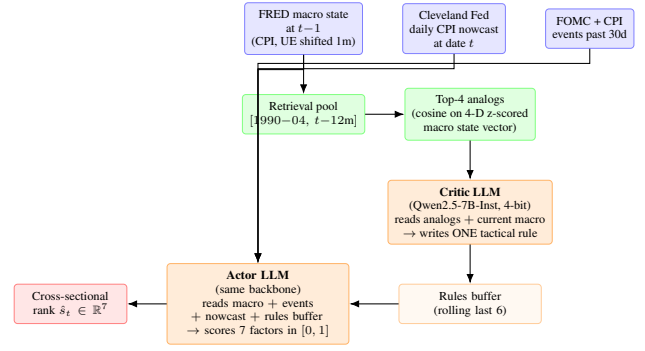
\begin{figure}[t]
\centering
\resizebox{\columnwidth}{!}{%
\begin{tikzpicture}[
  node distance=4mm and 8mm,
  every node/.style={font=\scriptsize},
  box/.style={draw, rounded corners=2pt, align=center, inner sep=4pt, minimum height=8mm},
  data/.style={box, fill=blue!10, draw=blue!50},
  llm/.style={box, fill=orange!15, draw=orange!60},
  retrieval/.style={box, fill=green!12, draw=green!50},
  output/.style={box, fill=red!10, draw=red!50},
  arrow/.style={-Latex, semithick}
]
\node[data] (fred) {FRED macro state\\at $t{-}1$\\(CPI, UE shifted 1m)};
\node[data, right=of fred] (cf) {Cleveland Fed\\daily CPI nowcast\\at date $t$};
\node[data, right=of cf] (events) {FOMC + CPI\\events past 30d};
\node[retrieval, below=8mm of fred] (pool) {Retrieval pool\\$[1990{-}04,\, t{-}12\text{m}]$};
\node[retrieval, right=of pool] (analogs) {Top-4 analogs\\(cosine on 4-D z-scored\\macro state vector)};
\node[llm, below=8mm of analogs] (critic) {\textbf{Critic LLM}\\(Qwen2.5-7B-Inst, 4-bit)\\reads analogs $+$ current macro\\$\to$ writes ONE tactical rule};
\node[box, fill=orange!5, draw=orange!40, below=8mm of critic, text width=2.6cm] (rules) {Rules buffer\\(rolling last 6)};
\node[llm, left=10mm of rules, text width=3.4cm] (actor) {\textbf{Actor LLM}\\(same backbone)\\reads macro $+$ events\\$+$ nowcast $+$ rules buffer\\$\to$ scores 7 factors in $[0,1]$};
\node[output, left=of actor, text width=2.0cm] (rank) {Cross-sectional\\rank $\hat{s}_t \in \mathbb{R}^7$};
\draw[arrow] (fred) -- (pool);
\draw[arrow] (pool) -- (analogs);
\draw[arrow] (analogs) -- (critic);
\draw[arrow] (critic) -- (rules);
\draw[arrow] (rules) -- (actor);
\draw[arrow] (fred.south) -- ++(0,-3mm) -| (actor.north);
\draw[arrow] (cf.south) -- ++(0,-3mm) -| (actor.north);
\draw[arrow] (events.south) -- ++(0,-3mm) -| (actor.north);
\draw[arrow] (actor) -- (rank);
\end{tikzpicture}
}
\caption{Leakage-controlled monthly decision pipeline. At each month-end $t$,
the system constructs only information observable at $t$: lag-shifted FRED
variables, archived Cleveland Fed CPI nowcasts, and recent macro-event
summaries. Historical analogs are retrieved from months at least 12 months
before $t$. A critic compresses the analog evidence into one rule; the actor
maps the current state and recent rules into seven factor scores.}
\label{fig:pipeline}
\end{figure}

\textbf{Architecture.} The pipeline is a two-LLM single-backbone actor-critic
with the critic conditioned on retrieved macro-analogous history~\citep{laskin2022context}
(Figure~\ref{fig:pipeline}). Both LLMs are Qwen2.5-7B-Instruct~\citep{qwen2_5_2024}
in 4-bit NF4 quantization via bitsandbytes~\citep{dettmers2023qlora}, greedy decoding ($T=0$);
month-by-month outputs are deterministic.

\textbf{Monthly decision protocol.} For each month-end $t$ we
\textit{(i)} construct a real-time macro state using lag-shifted FRED
variables and the archived CPI nowcast at $t$;
\textit{(ii)} retrieve top-$K$ analog months ($K=4$, matching the four
quadrants of the investment clock, \citealp{greetham2004investmentclock})
from $[1990\text{-}04,\, t-12\text{m}]$
by cosine similarity on the z-scored 4-D macro state vector $(\text{growth
pulse},\,\text{inflation pulse},\,\text{term spread},\,\text{VIX pulse})$,
chosen ex ante to mirror the investment-clock dimensions,
where pulses are 6-month changes and growth pulse $=-\Delta_6\text{(Unemployment,
shift 1)}$;
\textit{(iii)} ask the critic to compress analog evidence into one rule;
\textit{(iv)} append the rule to a rolling 6-rule buffer;
\textit{(v)} ask the actor to score seven factors in $[0,1]$;
\textit{(vi)} evaluate the Spearman rank IC against realised cross-sectional
ranks of market-adjusted residual factor returns. For every month $u$, we
standardise the macro state vector against the expanding history available
at $u$: $z_u = (x_u - \mu_{\le u}) / \sigma_{\le u}$. At decision month $t$,
cosine similarity is computed between $z_t$ and candidate $z_s$ for
$s \le t - 12$ months. No method uses a full-sample scaler. A representative
critic rule from 2023-04 (analogs: 1990-06, 2006-04, 2006-11, 2007-01) reads:
\emph{``Given the current low inflation and rising unemployment,
indicative of a growth-rising environment, raise HML by $0.15$ and
lower BAB by $0.15$, as historical analogs suggest HML tends to
outperform in such macro conditions.''}
The rule is parsed deterministically into a $\Delta$-vector applied to the
actor's seven-factor score before final ranking.

\textbf{Non-LLM baselines.} We compare the LLM pipeline against three
baselines drawn from the empirical asset pricing tradition~\citep{gu2020empirical}
that consume the same structured macro / nowcast information
set under the same decision-time constraint:
\textit{(i)} \emph{nowcast-only ridge}: per-factor Ridge regression with
$X = [\text{cf\_nowcast\_yoy},\,\text{cf\_nowcast\_delta\_6m}]$ only;
\textit{(ii)} \emph{macro + nowcast ridge}: per-factor Ridge on the full
6-feature decision-time vector (the four pulses plus the two nowcast columns),
with a \texttt{StandardScaler} fit on the training period only;
\textit{(iii)} \emph{kNN macro analog}: the same $K=4$ cosine retrieval as
the LLM pipeline, but instead of an LLM critic, output the per-factor mean
of the four analogs' realised next-month residual returns. Both Ridge models
train only on pre-test months with complete labels and decision-time features
($\approx 110$ months ending 2023-03; the effective start is set by the
Cleveland Fed nowcast archive coverage). The Ridge baselines are therefore
conservative relative to the kNN analog baseline, which pools macro states
from $[1990\text{-}04,\, t-12\text{m}]$ with the same $\geq 12$-month embargo
as the LLM pipeline and does not require the nowcast feature for analog
scoring. The principal apples-to-apples contrast in the results below is
therefore LLM-vs-kNN, both of which draw analogs from the same 33-year
history; the ridge baselines are reported as a sanity check on a smaller
training horizon.

\section{Results}\label{sec:results}

We report three checks. \textit{First}, does the LLM forecaster produce a
positive rank signal under decision-time inputs? \textit{Second}, is the
signal stable across non-overlapping contiguous sub-windows? \textit{Third},
does the gain come from the LLM architecture or from the real-time nowcast
input?

\begin{table}[t]
\centering
\caption{Monthly rank IC on the 36-month walk-forward (2023-04 to 2026-03).
All methods use decision-time inputs only. Macro+nowcast ridge has $n=34$
due to two missing unemployment observations; restricting all methods to
the common 34-month subset preserves the qualitative Full LLM--kNN
comparison: kNN remains comparable on median IC, while Full LLM remains
higher on mean IC.}
\label{tab:headline}
\scriptsize
\setlength{\tabcolsep}{3pt}
\begin{tabular}{lrrrr}
\toprule
Method & Mean & Median & 95\% CI & Perm.\ $p$ \\
\midrule
Base: zero-shot CoT             & $+0.007$ & $+0.027$ & $[-0.15, +0.16]$ & $0.92$ \\
\quad + raw analog retrieval    & $+0.006$ & $+0.063$ & $[-0.15, +0.16]$ & $0.94$ \\
\quad + critic rule synthesis   & $+0.088$ & $+0.009$ & $[-0.05, +0.22]$ & $0.21$ \\
Full LLM: \;+ CF nowcast        & $\mathbf{+0.131}$ & $+0.154$ & $[-0.02, +0.28]$ & $\mathbf{0.11}$ \\
\midrule
Nowcast-only ridge              & $+0.045$ & $+0.125$ & $[-0.14, +0.23]$ & $0.63$ \\
Macro + nowcast ridge           & $+0.058$ & $+0.071$ & $[-0.11, +0.23]$ & $0.51$ \\
kNN macro analog                & $+0.070$ & $\mathbf{+0.161}$ & $[-0.11, +0.24]$ & $0.44$ \\
\bottomrule
\end{tabular}
\end{table}

\begin{table}[t]
\centering
\caption{Three non-overlapping contiguous 12-month sub-windows of the full
LLM pipeline.}
\label{tab:subwindows}
\scriptsize
\setlength{\tabcolsep}{6pt}
\begin{tabular}{lr}
\toprule
Sub-window & Mean Spearman IC \\
\midrule
2023-04 to 2024-03 & $+0.111$ \\
2024-04 to 2025-03 & $+0.165$ \\
2025-04 to 2026-03 & $+0.117$ \\
\bottomrule
\end{tabular}
\end{table}

\textbf{Positive rank signal, but underpowered.} The full LLM pipeline
(Table~\ref{tab:headline}, row 4) achieves mean monthly rank IC $+0.131$ and
median $+0.154$, the strongest of the four LLM ablation rows. The bootstrap
95\% CI on the mean is $[-0.02, +0.28]$; the sign-symmetric permutation $p$
is $0.11$, which does not reach conventional significance.

\textbf{Sign-consistent across sub-windows.} Table~\ref{tab:subwindows} reports the
mean IC of the full pipeline on three non-overlapping contiguous 12-month
sub-windows; all three are positive ($+0.111, +0.165, +0.117$).

\textbf{What is doing the work?} The four-row ablation in
Table~\ref{tab:headline} isolates the contribution of each design element.
Zero-shot Chain-of-Thought (row 1) and zero-shot plus raw analog injection
into the actor's prompt (row 2) produce essentially the same mean and median
IC. The critic's rule synthesis (row 3) lifts the mean to $+0.088$ but
leaves the median near zero ($+0.009$). The single largest observed change
occurs when the Cleveland Fed CPI nowcast is added (row 4): the median
moves from $+0.009$ to $+0.154$.

\textbf{LLM versus non-LLM under the same decision-time constraint.}
The bottom three rows of Table~\ref{tab:headline} report ridge and kNN
baselines using the same structured macro / nowcast information set
under the same decision-time constraint. The kNN macro-analog baseline
recovers a slightly higher median IC than the full LLM pipeline
($+0.161$ vs $+0.154$). However, the kNN baseline's mean IC ($+0.070$) is
considerably below the LLM's ($+0.131$), and its permutation $p$ is
much higher ($0.44$ vs $0.11$). The two ridge baselines have positive
medians but mean IC below $+0.06$. We interpret this as: the macro-similar
retrieval step does much of the median-IC work, while the LLM pipeline
appears to improve the months in which the model commits most strongly to
top and bottom factors, which are the months that drive the long-short
sanity check.

\textbf{Economic sanity check.} Table~\ref{tab:econ} translates ranks into
a simple monthly allocation: long the top two factors and short the bottom
two, equal-weighted within each leg, with 5\,bps cost per unit weight
change. This is not a deployable trading strategy; it tests whether rank IC
corresponds to non-trivial allocation value.

\begin{table}[t]
\centering
\caption{Long-top-2 / short-bottom-2 sanity check, monthly rebalanced with
5\,bps per unit weight change. Turnover is the average monthly L1 weight
change (sum of $|w_t - w_{t-1}|$ across factors). \emph{Not a trading claim.}}
\label{tab:econ}
\scriptsize
\setlength{\tabcolsep}{4pt}
\begin{tabular}{lrrrr}
\toprule
Method & Ann.\ Ret. & Sharpe & Max DD & Turnover \\
\midrule
Full LLM pipeline       & $\mathbf{+4.2\%}$ & $\mathbf{+0.71}$ & $-4.7\%$  & $0.81$ \\
kNN macro analog        & $+1.5\%$ & $+0.29$ & $-5.1\%$  & $1.10$ \\
Nowcast-only ridge      & $-0.9\%$ & $-0.12$ & $-10.6\%$ & $0.17$ \\
Macro + nowcast ridge   & $-0.2\%$ & $-0.01$ & $-10.8\%$ & $0.82$ \\
\bottomrule
\end{tabular}
\end{table}

The LLM pipeline records the strongest Sharpe ($+0.71$) and annualised
return ($+4.2\%$); the kNN baseline is positive but lower ($+0.29$,
$+1.5\%$); the ridge baselines lose money over the test window. The
LLM-vs-kNN Sharpe gap is consistent with the mean-IC gap, suggesting that
the LLM's marginal contribution is concentrated in the extreme rankings that
drive long-short portfolio formation.

\section{Discussion}\label{sec:discussion}

\textbf{Main finding.} Our main finding is conditional and diagnostic. Under
a strict decision-time information constraint, a 7B retrieval-augmented LLM
produces a directionally positive factor-ranking signal on a 36-month
walk-forward, with median IC $+0.154$ and positive means across three
non-overlapping contiguous sub-windows. This magnitude is in the same range as
monthly cross-sectional IC reported for machine-learning factor models in
empirical asset pricing~\citep{gu2020empirical}. The largest observed change in
median IC occurs when the Cleveland Fed CPI nowcast is added. A simple kNN
macro-analog baseline (no LLM) recovers comparable median IC but materially
lower mean IC and lower long-short Sharpe. We therefore view the result as
two-layered: real-time inflation information and macro-similar retrieval
together account for much of the median signal, while the LLM appears to
improve the extreme rankings that drive the long-short sanity check.

\textbf{Why the nowcast matters.} The Cleveland Fed nowcast helps because it
substitutes for an unobservable quantity. At the month-end decision date for
month $t$, the BLS-published CPI for month $t$ does not yet exist, so any
backtest that conditions on it is consuming approximately ten days of
look-ahead about the realised macro state. Replacing that value with the
Cleveland Fed daily nowcast restores the information set that a real
decision-maker would have. This is a major and easily overlooked source of leakage we
identified, and matches the broader concern that financial backtests
routinely overstate performance through subtle look-ahead and selection
effects~\citep{harvey2020false,bailey2014deflated}. The kNN baseline benefits
from the same nowcast input, which is why it tracks the LLM on median IC;
the LLM's residual edge concentrates in extreme cross-sectional ranks rather
than in central tendency, suggesting the contribution is closer to
non-linear interaction across heterogeneous factor exposures than to
discovering a new signal de novo.

\textbf{Limitations.} \emph{Statistical power.} The evidence is underpowered: $n = 36$ monthly
decisions, mean-IC 95\% CI includes zero, permutation $p = 0.11$. With this
sample size and a per-month IC standard deviation near the Spearman null,
even a moderately positive expected IC fails to clear conventional
significance thresholds, and we explicitly avoid framing the result as a
discovery; the broader cross-sectional finance literature has documented how
fragile single-window evidence is once multiple-testing is taken into
account~\citep{harvey2016cross}. \emph{Information matching.} The LLM and kNN
paths are both leakage-clean but not perfectly information-matched---the LLM
critic anchors its macro narrative at $t{-}1$ and the actor additionally reads
recent event summaries (Appendix~\ref{app:repro})---so we read LLM-vs-kNN as a
comparison of logged decision protocols rather than a controlled input ablation.
\emph{Temporal dependence.} A moving-block bootstrap that accounts for
autocorrelation yields mean-IC 95\% CIs comparable to (and slightly tighter than)
the i.i.d.\ interval (Appendix~\ref{app:block-bootstrap}), so the rolling design
does not inflate the reported significance. \emph{Coverage.} The evaluation covers only U.S. equity
style factors and only one nowcast source. The benchmark's 2023-04 start is
set by the joint availability of the frozen nowcast archive, event corpus,
factor panel, and precomputed LLM decision logs under this pipeline
configuration. Extending the test window would require reconstructing a
longer frozen decision-time input archive and rerunning the walk-forward
pipeline. Two test months are unavailable for the \emph{macro + nowcast
ridge} baseline because its full six-feature vector is incomplete in our
frozen data snapshot. \emph{Deployability.} The economic sanity check is intentionally simple and
should not be interpreted as a deployable trading claim; deployable claims
require accounting for transaction costs, financing, and capacity, all of
which materially compress paper Sharpe ratios for academic
factors~\citep{frazzini2018trading}.

\textbf{Future work.} The natural extensions are (a) longer evaluation
windows once additional decision-time archive coverage is reconstructed, so
power against single-window tests is no longer the binding constraint, and
(b) explicit decomposition of the LLM's marginal contribution into the
extremes-vs-center dimension surfaced here, which would clarify whether the
mechanism is genuinely non-linear interaction or an artefact of small-$n$
ranking variance. Both directions can be pursued without changing the
leakage-controlled benchmark protocol introduced in this paper.

\textbf{Reproducibility.} We will release the walk-forward harness, all
prompt templates, the per-month score parquets for every method in
Tables~\ref{tab:headline}--\ref{tab:econ}, the parsed Cleveland Fed nowcast
archive (frozen snapshot 2026-05-13), the retrieved analog indices, the full
critic-rule log, and the exact Qwen2.5-7B-Instruct 4-bit (NF4 via
bitsandbytes) inference configuration with $T=0$ greedy decoding. All
reported bootstrap (10{,}000 resamples) and sign-symmetric permutation tests
(10{,}000) use \texttt{np.random.seed(42)}.

\bibliography{references}
\bibliographystyle{icml2024}

\onecolumn
\appendix

\section*{Appendix}

The appendix documents reproducibility metadata, leakage-control
details, a descriptive critic-rule audit, two representative case
studies, two compact sensitivity diagnostics, and extended figures.
It is descriptive and intended to make the system's behaviour
inspectable; it is not a causal claim about any individual factor's
performance in any individual month.

\section{Reproducibility and leakage controls}\label{app:repro}

We will release a public reproducibility package containing the
frozen input snapshots, per-method score files, retrieved analog
indices, critic-rule logs, prompt templates, and the analyses needed
to reproduce all tables and figures. The package includes the
Cleveland Fed nowcast snapshot frozen on 2026-05-13, the
Qwen2.5-7B-Instruct 4-bit NF4 inference configuration, greedy
decoding with $T=0$, and the random seeds used for bootstrap and
permutation tests (\texttt{seed} = 42).

All retrieval standardisation is expanding-window. For each month
$u$, we compute
\[
z_u \;=\; \frac{x_u - \mu_{\le u}}{\sigma_{\le u}},
\]
where $\mu_{\le u}$ and $\sigma_{\le u}$ are computed only from
information available at or before month $u$. Candidate months
satisfy $s \le t-12$ months, and no method uses a full-sample scaler.
We require at least 24 non-null monthly observations before computing
a standardised state.

The logged LLM protocol uses an as-of anchor $t-1$ for the
critic-side macro narrative, whereas the kNN baseline anchors
directly at $t$. Both anchors are leakage-clean because each path
only consumes information available at or before its anchor. We
therefore interpret LLM-vs-kNN comparisons as comparisons of the
logged decision protocol, not as a controlled sensitivity to the
standardisation anchor.

Prompt templates will be included in the public reproducibility
package.

\section{Critic-rule descriptive audit}\label{app:rule-audit}

We summarise the 36 critic outputs using deterministic keyword and
pattern rules; no LLM-as-judge is used. Table~\ref{tab:rule-themes}
shows that the rules are highly concentrated in this window: of the
30 rules with extractable factor pairs, 25 raise HML, usually under
inverted-yield-curve / low-volatility tags. This supports two
readings. First, the LLM behaviour is inspectable rather than
arbitrary. Second, the 2023--2026 window is regime-specific, so the
rule distribution should not be interpreted as evidence of general
factor-timing skill.

\begin{table}[h]
\centering
\small
\caption{Recurring critic-rule themes on the 36 monthly decisions of
the test window. Macro regime tags are abbreviated: \texttt{iyc} $=$
inverted yield curve, \texttt{lv} $=$ low volatility. Total count of
categorised (raise, lower) pairs $=30$; the remaining 6 rules have no
extractable factor pair.}
\label{tab:rule-themes}
\begin{tabular}{lll r}
\toprule
Macro regime tag(s) & Raise & Lower & Count \\
\midrule
iyc, lv & HML & BAB & 7 \\
iyc, lv & HML & UMD & 6 \\
iyc, lv & HML & CMA & 5 \\
iyc     & HML & UMD & 4 \\
iyc, lv & SMB & UMD & 2 \\
iyc, lv & QMJ & BAB & 1 \\
iyc, lv & HML & RMW & 1 \\
iyc     & UMD & QMJ & 1 \\
iyc     & HML & QMJ & 1 \\
iyc     & UMD & BAB & 1 \\
other / mixed regime tags & HML & BAB & 1 \\
\bottomrule
\end{tabular}
\end{table}

\section{Case studies}\label{app:case-studies}

We present two pre-selected decision months: the best single-month
Full LLM IC, and the month with the largest Full-LLM minus kNN IC
gap. These examples are illustrative, not causal; they show how the
logged critic rule and actor scores relate to the realised rank
ordering in two months where the LLM and kNN paths diverge. The critic
rules quoted below are reproduced verbatim from the model output; any
citation-like text within them (e.g., ``Asness-Liew-Pedersen 2013'') is
generated by the model and is not an author citation.

\subsection{Best LLM month: 2024-12}

\textbf{Decision date.} 2024-12-31.
\textbf{Decision-time macro snapshot.} VIX 17.4; term spread $+0.33$;
HY credit spread $2.92$; real 10y yield $2.24$; shift-1 unemployment
$4.10$; shift-1 FRED CPI YoY $2.87$; Cleveland Fed CPI YoY nowcast at
$t$: $\mathbf{2.86}$.

\textbf{Retrieved analogs ($K=4$).}
1990-06 ($\rho=+0.96$); 2019-03 ($\rho=+0.96$); 2023-10 ($\rho=+0.99$);
2023-12 ($\rho=+0.98$).

\textbf{Critic.}
\textit{Rule:} ``Given the inverted yield curve and low equity
volatility, raise the score of HML by 0.15 and lower the score of
UMD by 0.15, as historical analogs indicate that value factors tend
to outperform in such macro conditions (Asness-Liew-Pedersen 2013).''
\textit{Diagnosis:} ``In all analogs, HML consistently wins while UMD
consistently loses.''

\textbf{Actor scores vs realised ranks.}
\begin{tabular}{l|rrrrrrr}
Factor    & SMB & HML & RMW & CMA & UMD & BAB & QMJ \\
\midrule
Score     & 0.20 & 0.30 & 0.80 & 0.40 & 0.35 & 0.50 & 0.70 \\
Pred rank & 1 & 2 & 7 & 4 & 3 & 5 & 6 \\
Realised  & 1 & 2 & 7 & 3 & 4 & 5 & 6 \\
\end{tabular}

\textbf{Outcome.} Full LLM IC $= +0.964$; kNN IC $= -0.321$;
gap $= +1.286$.

\textit{Reading.} December 2024 sits in the persistent
inverted-yield-curve / low-volatility regime that characterises most
of the test window. The retrieved analogs span 1990, 2019, and two
2023 months, and the critic converges on a value-tilted prior. This
case is consistent with the actor recovering the realised
cross-sectional ordering on five of seven factors and swapping only
two adjacent middle factors (CMA, UMD); the kNN baseline mis-ranks
top and bottom factors in the same month.

\subsection{Largest LLM-beats-kNN gap: 2024-07}

\textbf{Decision date.} 2024-07-31.
\textbf{Decision-time macro snapshot.} VIX 16.4; term spread $-0.20$;
HY credit spread $3.25$; real 10y yield $1.85$; shift-1 unemployment
$4.20$; shift-1 FRED CPI YoY $2.94$; Cleveland Fed CPI YoY nowcast at
$t$: $\mathbf{3.01}$.

\textbf{Retrieved analogs ($K=4$).}
1990-05 ($\rho=+0.99$); 2006-03 ($\rho=+0.94$); 2006-08 ($\rho=+0.99$);
2006-09 ($\rho=+0.94$).

\textbf{Critic.}
\textit{Rule:} ``Given the inverted yield curve and low equity
volatility, raise the score of HML by 0.15 and lower the score of CMA
by 0.15, as historical analogs indicate that value factors tend to
outperform in such macro conditions (Asness-Liew-Pedersen 2013).''
\textit{Diagnosis:} ``In all analogs, HML consistently wins while
CMA and QMJ consistently lose.''

\textbf{Actor scores vs realised ranks.}
\begin{tabular}{l|rrrrrrr}
Factor    & SMB & HML & RMW & CMA & UMD & BAB & QMJ \\
\midrule
Score     & 0.70 & 0.80 & 0.30 & 0.40 & 0.60 & 0.50 & 0.20 \\
Pred rank & 6 & 7 & 2 & 3 & 5 & 4 & 1 \\
Realised  & 7 & 6 & 3 & 4 & 1 & 5 & 2 \\
\end{tabular}

\textbf{Outcome.} Full LLM IC $= +0.607$; kNN IC $= -0.857$;
gap $= +1.464$.

\textit{Reading.} July 2024 has a similar macro signature to 2024-12
but a different mix of analogs: three of four come from pre-GFC 2006
episodes. The critic again produces a value-tilted prior. The actor
gets the broad direction right but mis-orders UMD. This case is
consistent with the critic rule acting as a coarse value-tilted
abstraction of the analog evidence rather than a mechanical average
of analog returns.

\section{Sensitivity and extreme-rank diagnostics}\label{app:diagnostics}

\subsection{Sensitivity to analog count \(K\)}\label{app:k-sensitivity}

We vary \(K\) only in the LLM-stripped kNN macro-analog baseline over
\(K \in \{2,4,6,8,12\}\), keeping the retrieval features, embargo,
labels, and long-short allocation rule fixed. The LLM actor-critic
pipeline is not rerun. This diagnostic tests whether the retrieval
signal is specific to the \(K=4\) choice used in the main experiment.

\begin{table}[h]
\centering
\small
\caption{kNN macro-analog sensitivity to the number of retrieved
analogs \(K\).}
\label{tab:k-sensitivity}
\begin{tabular}{rrrr}
\toprule
\(K\) & Mean IC & Median IC & Long-short Sharpe \\
\midrule
2  & $+0.017$ & $+0.089$ & $+0.02$ \\
4  & $+0.070$ & $+0.161$ & $+0.29$ \\
6  & $+0.086$ & $+0.089$ & $+0.51$ \\
8  & $+0.133$ & $+0.179$ & $+0.77$ \\
12 & $+0.079$ & $+0.089$ & $+0.17$ \\
\bottomrule
\end{tabular}
\end{table}

The kNN signal is not unique to \(K=4\): several nearby choices
remain positive, and \(K=8\) is strongest in this diagnostic. We
retain \(K=4\) in the main comparison because it was fixed ex ante
to match the LLM retrieval protocol; the sweep is reported only as a
sensitivity check.

\subsection{Top/bottom overlap}\label{app:extreme-overlap}

Because the long-short sanity check depends only on the top and
bottom of the predicted factor ranking, we report top-2 and bottom-2
overlap with realised factor ranks for each decision month and
average across the 36-month window. This diagnostic makes the
extreme-rank interpretation directly observable; it is not a
separate trading claim and does not prove LLM superiority.

We define:
\[
\mathrm{Top2Overlap}_t \;=\;
  \frac{\bigl|\mathrm{Top2}_{\mathrm{pred},t} \cap \mathrm{Top2}_{\mathrm{real},t}\bigr|}{2},
\quad
\mathrm{Bottom2Overlap}_t \;=\;
  \frac{\bigl|\mathrm{Bottom2}_{\mathrm{pred},t} \cap \mathrm{Bottom2}_{\mathrm{real},t}\bigr|}{2}.
\]
\[
\mathrm{ExtremeOverlap}_t \;=\;
  \tfrac{1}{2}\bigl(\mathrm{Top2Overlap}_t + \mathrm{Bottom2Overlap}_t\bigr).
\]

\begin{table}[h]
\centering
\small
\caption{Extreme-rank overlap diagnostics, averaged over the
36-month walk-forward (34 months for the macro + nowcast ridge).}
\label{tab:extreme-overlap}
\begin{tabular}{lrrr}
\toprule
Method & Top-2 & Bottom-2 & Extreme \\
\midrule
Full LLM pipeline       & $+0.417$ & $+0.278$ & $+0.347$ \\
kNN macro analog        & $+0.347$ & $+0.361$ & $+0.354$ \\
Nowcast-only ridge      & $+0.278$ & $+0.389$ & $+0.333$ \\
Macro + nowcast ridge   & $+0.309$ & $+0.353$ & $+0.331$ \\
\bottomrule
\end{tabular}
\end{table}

The overlap diagnostic is mixed: Full LLM has the highest Top-2
overlap, whereas kNN has slightly higher Bottom-2 and average
Extreme overlap. We therefore treat this table as an interpretability
diagnostic rather than evidence of uniformly better extreme-rank
recovery; Table~3 remains the allocation-level sanity check.

\subsection{Robustness to temporal dependence}\label{app:block-bootstrap}

The rolling walk-forward induces temporal dependence in the monthly IC
series that an i.i.d.\ bootstrap ignores. We recompute the full-pipeline
mean-IC 95\% confidence interval with a moving-block bootstrap
(10{,}000 resamples, seed $42$) at block lengths $L \in \{3, 6\}$ months.
The resulting intervals are $[+0.013, +0.268]$ ($L{=}3$) and
$[+0.039, +0.228]$ ($L{=}6$), comparable to and slightly tighter than the
i.i.d.\ interval $[-0.023, +0.282]$ reported in Table~\ref{tab:headline}.
The monthly IC series has mildly negative lag-1 autocorrelation ($-0.14$),
so accounting for temporal dependence does not widen the interval. We
nonetheless treat the mean-IC evidence cautiously given $n=36$ and the
sign-symmetric permutation $p=0.11$; this check addresses whether the rolling
design's serial dependence inflates the reported significance.

\section{Extended figures}\label{app:figures}

\begin{figure}[H]
\centering
\includegraphics[width=0.95\linewidth]{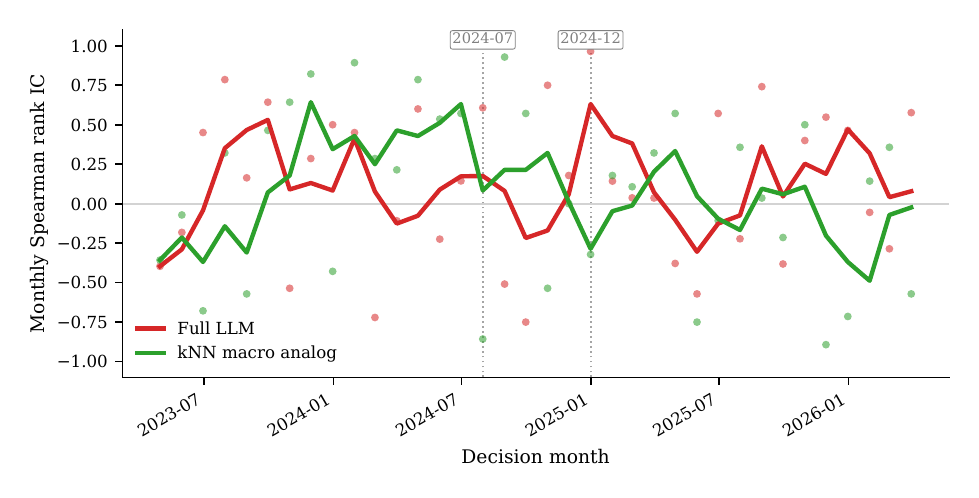}
\caption{Monthly rank IC for the two main analog-based methods.
Points show raw monthly IC; thick lines show 3-month rolling averages
for readability. Summary statistics in Table~1 use unsmoothed
monthly IC. Marked months correspond to the case studies in
Appendix~\ref{app:case-studies}.}
\label{fig:ic-ts}
\end{figure}

\begin{figure}[H]
\centering
\includegraphics[width=0.95\linewidth]{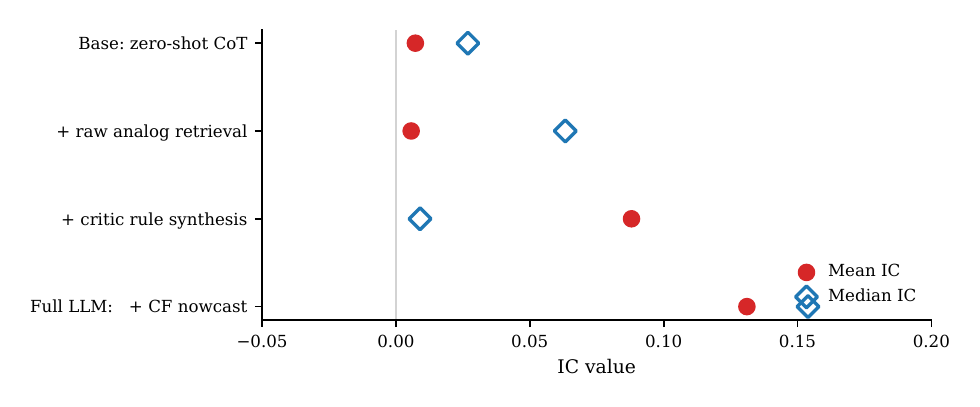}
\caption{LLM ablation summary. Points show mean and median monthly
rank IC for the four ablation rows. The largest observed median-IC
change occurs when the Cleveland Fed nowcast is added.}
\label{fig:ic-ts-ablation}
\end{figure}

\begin{figure}[H]
\centering
\includegraphics[width=0.95\linewidth]{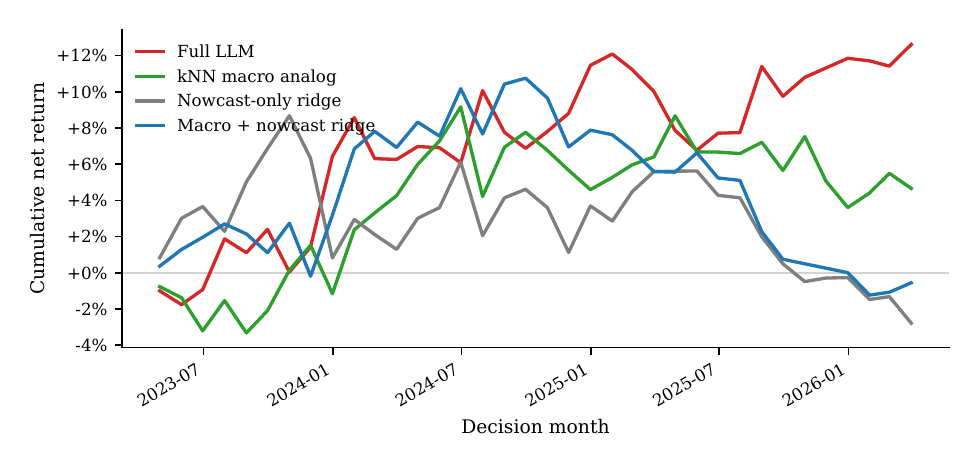}
\caption{Cumulative net return of the long-top-2 / short-bottom-2
sanity check (monthly rebalanced, 5\,bps per unit weight change),
following the methodology of Table~3 of the main paper. This plot
visualises the allocation sanity check only and is not evidence of
deployable trading performance.}
\label{fig:cumret}
\end{figure}

\end{document}